\documentclass{sf2a-conf2013}
\usepackage{graphicx}
\usepackage{hyperref}
\usepackage[]{natbib}  
\usepackage{epstopdf}

\def\NIKA{\textit{NIKA}}
\def\NIKAII{\textit{NIKA2}}
\def\BibTeX{{\rm B\kern-.05em{\sc i\kern-.025em b}\kern-.08em
    T\kern-.1667em\lower.7ex\hbox{E}\kern-.125emX}}
\bibpunct{(}{)}{;}{a}{}{,}  

\begin{document}

\TitreGlobal{SF2A 2013}


\title{Detection of the tSZ effect with the $\NIKA$ camera}

\runningtitle{tSZ imaging with NIKA}

\author{B.~Comis}
\address{Laboratoire de Physique Subatomique et de Cosmologie,
Universit\'e Joseph Fourier Grenoble 1,
  CNRS/IN2P3, Institut Polytechnique de Grenoble, 
  53, rue des Martyrs, Grenoble, France}
  
\author{R.~Adam$^1$}

\author{J.F.~Mac\'ias-P\'erez$^1$}

\author{A.~Adane}
\address{Institut de RadioAstronomie Millim\'etrique (IRAM), Grenoble, France}

\author{P.~Ade}
\address{Astronomy Instrumentation Group, University of Cardiff, UK}

\author{P.~Andr\'e}
\address{Laboratoire AIM, CEA/IRFU, CNRS/INSU, Universit\'e Paris Diderot, CEA-Saclay, 91191 Gif-Sur-Yvette, France}

\author{A.~Beelen}
\address{Institut d'Astrophysique Spatiale (IAS), CNRS and Universit\'e Paris Sud, Orsay, France}

\author{B.~Belier}
\address{Institut d'Electronique Fondamentale (IEF), Universit\'e Paris Sud, Orsay, France}

\author{A.~Beno\^it}
\address{Institut N\'eel, CNRS and Universit\'e de Grenoble, France}

\author{A.~Bideaud$^3$}
\author{N.~Billot}
\address{Institut de RadioAstronomie Millimetrique (IRAM), Granada, Spain}

\author{O.~Bourrion$^1$}
\author{M.~Calvo$^7$}
\author{A.~Catalano$^1$}
\author{G.~Coiffard$^2$}
\author{A.~D'Addabbo$^7$}
\author{F.-X.~D\'esert}
\address{Institut de Plan\'etologie et d'Astrophysique de Grenoble (IPAG), CNRS and Universit\'e de
Grenoble, France}

\author{S.~Doyle$^3$}
\author{J.~Goupy$^7$}
\author{C.~Kramer$^8$}
\author{S.~Leclercq$^2$}
\author{J.~Martino$^5$}
\author{P.~Mauskopf$^3$}
\author{F.~Mayet$^1$}
\author{A.~Monfardini$^7$}
\author{F.~Pajot$^5$}
\author{E.~Pascale$^3$}
\author{E.~Pointecouteau$^{10, }$}
\address{Universit\'e de Toulouse; UPS-OMP; IRAP;  Toulouse, France}
\address{CNRS; IRAP; 9 Av. colonel Roche, BP 44346, F-31028 Toulouse cedex 4, France}
\author{N.~Ponthieu$^9$}
\author{V.~Rev\'eret$^4$}
\author{L.~Rodriguez$^4$}
\author{G.~Savini $^{4,}$}
\address{University College London, Department of Physics and Astronomy, Gower Street, London WC1E 6BT, UK}

\author{K.~Schuster$^2$}
\author{A.~Sievers$^8$}
\author{C.~Tucker$^3$}
\author{R.~Zylka$^2$}

\setcounter{page}{1}


\maketitle

\begin{abstract}
We present the first detection of the thermal Sunyaev-ZelÕdovich (tSZ) effect from a cluster of galaxies performed with a KIDs (Kinetic Inductance Detectors) based instrument. The tSZ effect is a distortion of the black body CMB (Cosmic Microwave Background) spectrum produced by the inverse Compton interaction of CMB photons with the hot electrons of the ionized intra-cluster medium.

The massive, intermediate redshift cluster RX J1347.5-1145 has been observed using {\it NIKA} (New IRAM KIDs arrays), a dual-band (140 and 240 GHz) mm-wave imaging camera, which exploits two arrays of hundreds of KIDs: the resonant frequencies of the superconducting resonators are shifted by mm-wave photons absorption.

This tSZ cluster observation demonstrates the potential of the next generation {\it NIKA2} instrument\footnote{\url{http://ipag.osug.fr/nika2}}, being developed for the 30m telescope of IRAM, at Pico Veleta (Spain). {\it NIKA2} will have 1000 detectors at 140GHz and 2x2000 detectors at 240GHz, providing in that band also a measurement of the linear polarization. {\it NIKA2} will be commissioned in 2015.
\end{abstract}



\section{Introduction}
The important role of millimeter wave astronomy for both cosmology and astrophysics is today well established. 
However, the standard technology for these wavelengths, low-temperature bolometers, requires complex cryogenic readout electronics, which limits the reachable multiplexing ratios well below a hundred. In order to increase the focal plane area and pixel number, an alternative technology intrinsically suited for frequency domain multiplexing, able to provide multiplexing ratios greater than a thousand, looks extremely interesting. It is therefore worth studying and employing Kinetic Inductance Detectors (KIDs) as a promising alternative to bolometers.

The New IRAM KIDs Array ($\NIKA$) is a dual-band (140 and 240 GHz) KIDs camera developed in Grenoble to work at the IRAM 30m telescope. The first four $\NIKA$ commissioning campaigns at the IRAM 30 m telescope  \citep{Monfardini2010, Monfardini2011, Calvo2013} have allowed us to demonstrate performances comparable to state-of-the-art bolometer arrays, working at the same wavelengths \citep[e.g. GISMO,][observing only at 2 mm]{Staguhn2008}.

We report here the first KID-based observation of a galaxy cluster via the thermal Sunyaev-ZelÕdovich (tSZ) effect, RX J1347.5-1145 has been imaged using the NIKA prototype \citep{Adam2013} during the fifth observing run \citep[November 2012,][]{main_run5}. When observed through the tSZ effect, clusters of galaxies show up as weak extended sources. In fact, this effect is a small spectral distortion (of the order of one-thousandth) of the cosmic microwave background (CMB) intensity, and its detection and mapping have been hard challenges. However, recently, tSZ-selected cluster catalogues containing hundreds of candidates with arcmin resolution have finally been produced (SPT -- \citealp{SPT_cat}, ACT -- \citealp{Hasselfield} and the Planck Satellite --  \citealp{PSZ_cat}). Consequently, high-resolution tSZ observations and follow-ups are now necessary to deeply explore the cluster internal structure. The NIKA camera at the IRAM 30 m telescope is a well-suited instrument for such observations and follow-ups, given its resolution, its sensitivity and the two observing frequencies.

\section{The $\NIKA$ camera}

Located on Pico Veleta, in a dry and high (2850 m a.s.l.) area, the IRAM 30-m telescope is one of today's largest and most sensitive for mm wavelenghts. In order to completely exploit its angular resolution and the entire field of view, large arrays of (thousands) detectors are needed. For this reason it is worth to develop a camera based on an alternative technology with respect to bolometers, the Kinetic Inductance Detectors. KIDs are superconducting resonators whose resonance frequency changes with the absorbed optical power. The resonant frequencies of the individual resonators can be easily controlled geometrically during array design \citep{Monfardini2010} and  a very large number of pixels, of the order of thousands, can be multiplexed on a single transmission line. Compared to bolometers, KIDs also offer further advantages. 
They are significantly less sensitive to temperature fluctuations, microphonic noise and magnetic field fluctuations (with respect to Transition Edge Sensor Bolometers).

The New IRAM KIDs Array ($\NIKA$) is a dual-band KIDs camera custom designed for the IRAM 30-m Nasmyth-focus telescope, whose detectors are cooled down to $\sim$ 100 mK with a $^4$He -- $^3$He dilution cryostat.
$\NIKA$, a prototype of the final camera $\NIKAII$,  consists of two arrays of Kinetic Inductance Detectors (KIDs) with maximum transmission at about 140 and 240 GHz and respective angular resolutions (FWHM) of 18.5 and 12.5 arcsec and effective fields of view of 1.8 $\times$1.8 and 1.0 $\times$ 1.0 arcmin. During the test run of November 2012, the 140 GHz band was used with 127 detectors with a mean effective sensitivity of 29 mJy.s$^{1/2}$ per beam. Concerning the 240 GHz channel, despite the disfunction of a cold amplifier during this campaign, 91 detectors with mean effective sensitivity of 55 mJy.s$^{1/2}$ per beam  where used, recovering the expected mean effective sensitivity of 37 mJy.s$^{1/2}$ per beam by using only eight detectors. 
More details on the NIKA prototype setup (Run5) can be found in NIKA Collaboration (2013).
  
\section{First tSZ results: mapping of RX J1347.5-1145} 

	\begin{figure*}	
	\centering	
	\includegraphics[width=0.3\textwidth]{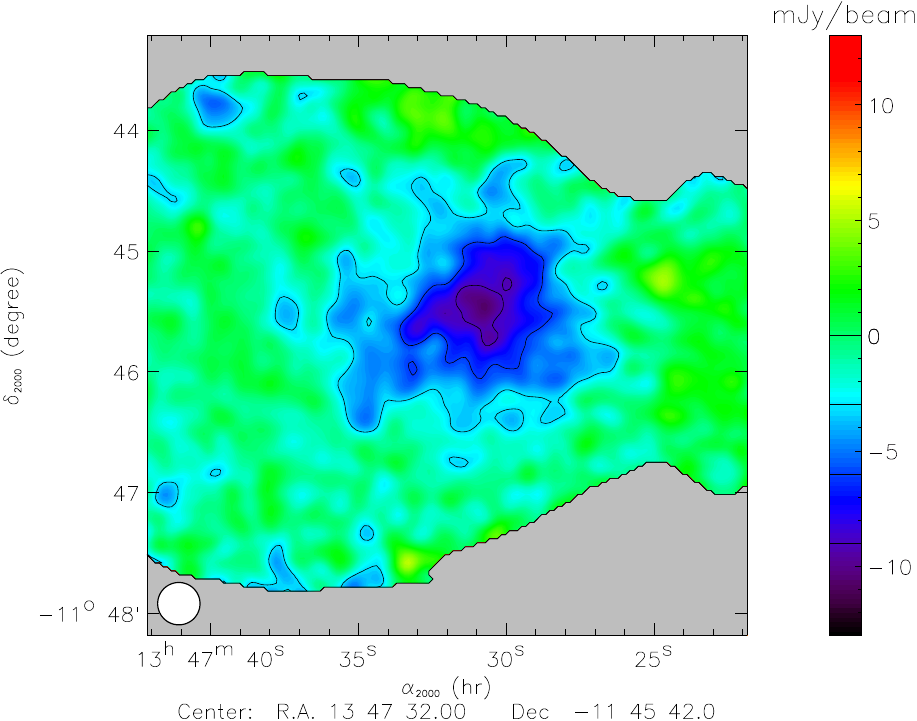}
	\hspace*{0.3cm}
	\includegraphics[width=0.3\textwidth]{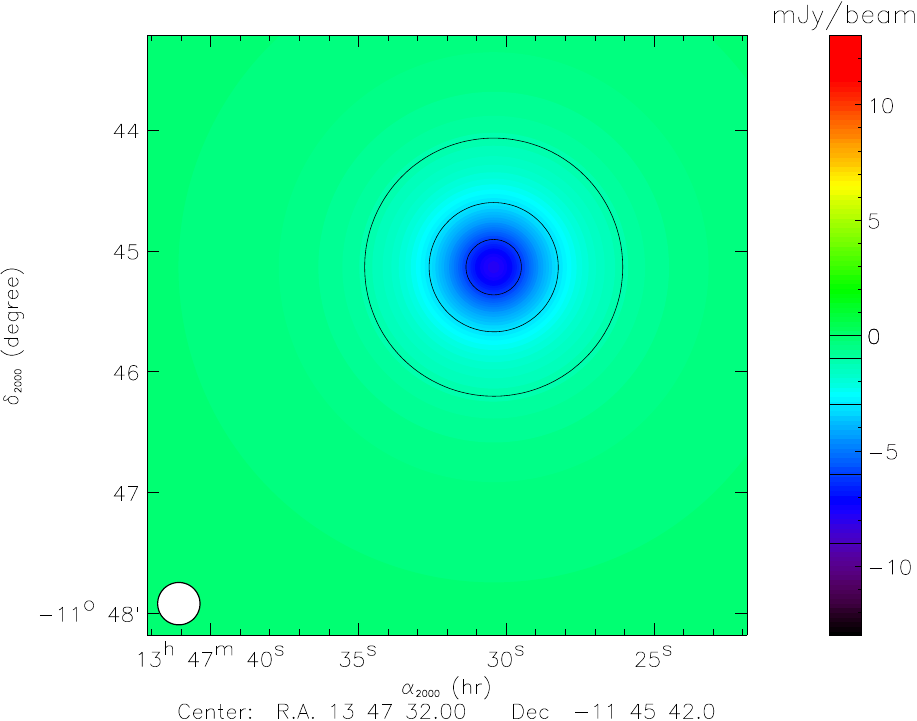}
	\hspace*{0.3cm}
	\includegraphics[width=0.3\textwidth]{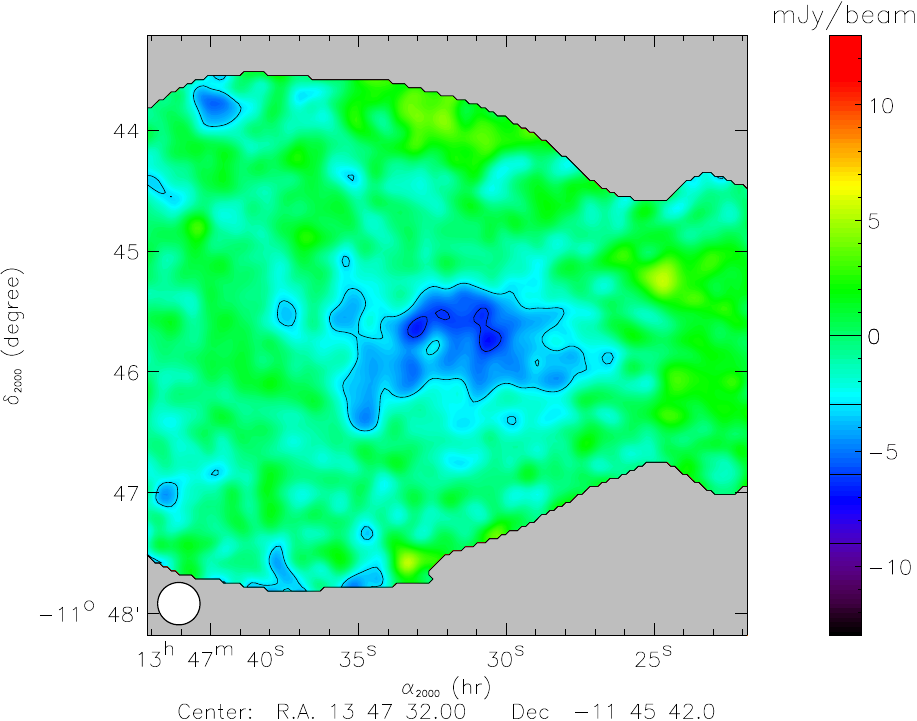}
	\caption{{\bf Left:} RX~J1347.5-1145 reconstructed tSZ map (at 140 GHz), the central radio point source (4.4 mJy) has been subtracted. {\bf Middle:} Map of the best fitting model for the relaxed cool-core like region, obtained by masking the shock area. {\bf Right:} Residuals obtained by subtracting the fitted model to the original map. As expected, the model accounts well for the cluster profile except in the southern shocked area.}
        \label{fig:RXJ1347}
	\end{figure*}

\begin{figure}[ht!]
 \centering   
 \includegraphics[width=0.48\textwidth,clip]{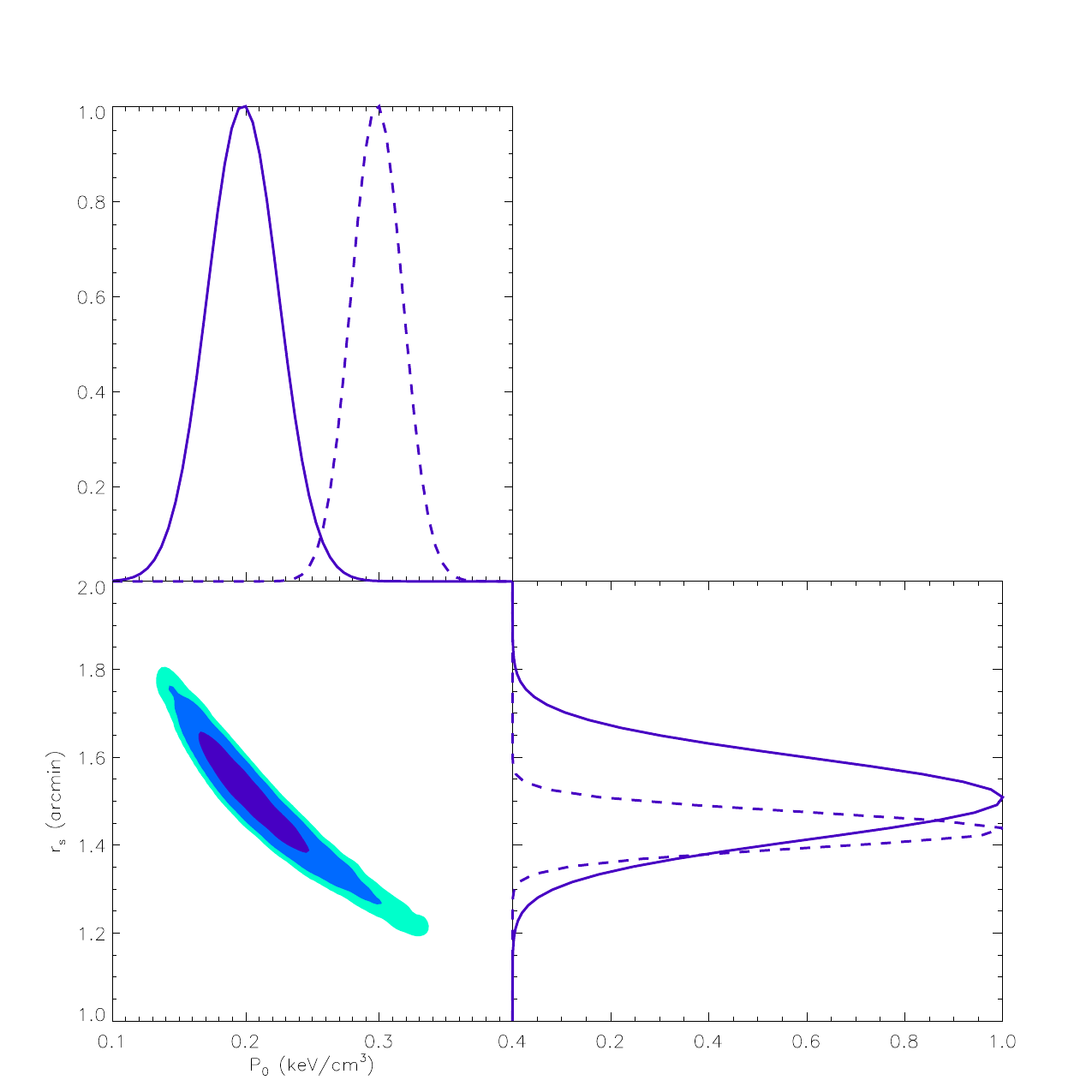}      
 \caption{Posterior likelihood of the MCMC pressure profile fit in the plane $P_0$ -- $r_s$. From dark to light blue, the colors correspond to 68.2\%, 95.4\% and 99.7\% confidence levels. The top and right curves show the marginalized normalized likelihood of $P_0$ and $r_s$ respectively. The dashed lines allow comparison with the marginalized distribution obtained when no mask is applied to the map.}
  \label{fig:likelihood}
\end{figure}

Clusters of galaxies are the largest gravitationally bound structures observable in our Universe.
They are mainly ($\sim$ 80\%) constituted of dark matter, while most of the baryons are in the form of a hot ionized gas, the so-called intra-cluster medium (ICM). The ICM is responsible for the Bremsstrahlung X-ray emission but also for the tSZ effect, a distortion of the black body CMB spectrum produced by the inverse Compton interaction of CMB photons with the hot electrons of the ICM \citep{SunyaevZeldovich1, SunyaevZeldovich2}. This interaction produces a unique spectral signature, a decreased CMB intensity at frequencies lower than $\sim$217 GHz, and an increase at higher frequencies, making the NIKA channels particularly interesting for SZ studies. While the frequency behavior is specific of the effect, the amplitude of the signal is defined by the Comptonization parameter
\begin{equation}
	y = \frac{\sigma_{\mathrm{T}}}{m_{\mathrm{e}} c^2} \int P_{\mathrm{e}} dl
	\label{eq:y_compton}
\end{equation}
(with $m_e$ the electron rest mass, $c$ the light speed, $\sigma_T$ the electron Thomson scattering cross-section), which is then a measure of the integrated electronic pressure $P_{\mathrm{e}}$ along the line-of-sight, $\vec{l}$.

RX J1347.5-1145 is one of the most extensively studied clusters. Located at intermediate redshift ($z$ = 0.451), it represented a particularly well-suited candidate for the $\NIKA$ 5$^{th}$ run, being both compact and strong enough for the prototype f.o.v. and sensitivity, as estimated by exploiting the best fitting $P_e(r)$ obtained for the work of \cite{Comis2011} from the X-ray derived pressure profiles of the $ACCEPT$ catalogue \citep{Cavagnolo2009}.
Furthermore, this cluster is a perfect illustration of the complementarity of tSZ effect and X-ray signals: initially thought to be a well-relaxed (cool-core) object according to its first X-ray data \citep{Schindler1997},  later tSZ observations showed a substructure located at $\sim$ 20" from the center towards the SE region \citep{Komatsu1999, Pointecouteau1999}. This has been interpreted as a hotter, over-pressured component resulting from a merging event. The non-trivial morphology makes this cluster an ideal target to test $\NIKA$ capabilities of probing the details of ICM physics.

In Fig.~\ref{fig:RXJ1347} left panel) we report the 140 GHz tSZ map obtained during the $\NIKA$ 5$^{th}$ run, with a total (unflagged) observing time of 5h 47min (scans of 6min 20s). The analysis used 81 detectors at 140 GHz and 45 at 240 GHz (for the dual-band decorrelation of the atmospheric noise, further details in \citealp{Adam2013}). 

In order to extract the signal from the shock, produced by the ongoing merger, we have modeled the relaxed region by considering a generalized Navarro, Frenk and White \citep[gNFW,][]{gNFW} pressure profile
\begin{equation}
	P(r) = \frac{P_0}{\left(\frac{r}{r_{\mathrm{s}}}\right)^{\gamma}\left[1+\left(\frac{r}{r_{\mathrm{s}}}\right)^{\alpha}\right]^{\frac{\beta-\gamma}{\alpha}}},
\end{equation} 
centered at the X-ray position of the system. $P_0$ is a scalar normalization of the pressure profile, while the parameters $\alpha$, $\beta$, $\gamma$ are respectively the slopes for intermediate radii ($r \sim r_s$, a scale radius), the outer region ($r >> r_s$) and the core region ($r << r_s$).  The intermediate, outer and inner slopes have been set equal to the cool-core best fitting values of \cite{Arnaud2010}, ($\gamma$ = 0.7736, $\beta$ = 5.4905, $\alpha$ = 1.2223).
The free parameters $P_0$ and $r_s$ have been constrained through a Markov
Chain Monte Carlo (MCMC) approach, masking the South East extension with a half ring, centered on the X-ray peak, with inner and outer radii set to 10 and 80 arcsec. The values obtained are
	$P_{0}  =  0.198 \pm 0.028   \pm^{0.054}_{0.037} \mathrm{keV/cm}^3$ and 
	$r_s  =   1.510 \pm 0.093  \pm^{0.30}_{0.00}  \mathrm{arcmin}$.
The systematic uncertainties have been computed using the calibration uncertainty and taking into account bias filtering effects of the analysis estimated from the difference in the input and output simulated profiles \citep{Adam2013}.
Fig.~\ref{fig:RXJ1347} allows comparison between the $\NIKA$ prototype tSZ map (left panel), the best fitting model obtained for the relaxed component (middle panel), and the residual (right panel). The model represents well the northern part of the tSZ map but the southern side cannot be explained without including an overpressure component, that is known to be due to the merging of a sub-cluster \citep{Cohen2002, Allen2002, Miranda2008}.
In Fig.~\ref{fig:likelihood} the posterior likelihood of the model parameters is given. Note that in the case of no mask being applied to the map, $r_s$ is still compatible with the value given above (Fig.~\ref{fig:likelihood}, dashed line).

\section{Conclusions}
Past NIKA runs have demonstrated the potential for KIDs to provide large pixel arrays for ground-based millimeter astronomy. Now we have also shown the tSZ capabilities of a KIDs-based camera.

The agreement with DIABOLO observations of RX J1347.5Ð1145 \citep{Pointecouteau1999, Pointecouteau2001}, performed at the same telescope and with similar resolution and frequency coverage, has been used to validate the $\NIKA$ tSZ map \citep{Adam2013}. In addition, the $\NIKA$ prototype map agrees also with state-of-the-art sub-arcminute resolution tSZ observations, MUSTANG \citep[90 GHz and 8 arcsec resolution]{Korngut2011} and $CARMA$ \citep[30 Ð 90 GHz and $\sim$15 arcsec resolution]{Plagge2012}.

$\NIKA$ 5$^{th}$ run has proven that KID arrays are competitive detectors for millimeter wave astronomy and in particular for the observation of galaxy clusters via the tSZ effect. The prototype $\NIKA$ is now fully operational and will be open to the science community strating from the next Winter (2013/14). 
The second generation instrument, $\NIKAII$ (\url{http://ipag.osug.fr/nika2}), will be made of about 1000 detectors at 140 GHz and 4000 at 240 GHz with a field of view of $\sim$ 6.5 arcmin. With these characteristics, $\NIKAII$ will be able to provide larger f.o.v.,  high resolution mapping of clusters. Therefore a perfect instrument for high resolution observations and follow-ups of medium and high redshift cluster.  $\NIKAII$ will be installed for commissioning in 2015. 

\begin{acknowledgements}
\textbf{Acknowledgements:} This work has been partially funded by the Foundation
Nanoscience Grenoble, the ANR under the contracts ÓMKIDSÓ and ÓNIKAÓ.
This work has been partially supported by the LabEx FOCUS ANR-11-LABX-
0013. This work has benefited from the support of the European Research
Council Advanced Grant ORISTARS under the European UnionÕs Seventh
Framework Programme (Grant Agreement no. 291294). 
The NIKA dilution cryostat has been designed and built by the Institut N\'eel Cryogenics Group. In particular, we acknowledge the crucial contributions of Gregory Garde, Henri Rodenas, Jean-Paul Leggeri, Philippe Camus.
EP acknowledges the support of grant ANR-11-BS56-015. R. A. would like to thank the ENIGMASS French LabEx for funding this work, B. C. acknowledges support from the CNES post-doctoral fellowship program. 
\end{acknowledgements}

\bibliographystyle{aa} 
\bibliography{sf2a_NIKA_tSZ}

\begin{thebibliography}{24}
\expandafter\ifx\csname natexlab\endcsname\relax\def\natexlab#1{#1}\fi

\bibitem[{{Adam} {et~al.}(2013){Adam}, {Comis}, {Mac{\'{\i}}as-P{\'e}rez},
  {Adane}, {Ade}, {Andr{\'e}}, {Beelen}, {Belier}, {Beno{\^i}t}, {Bideaud},
  {Billot}, {Boudou}, {Bourrion}, {Calvo}, {Catalano}, {Coiffard}, {D'Addabbo},
  {D{\'e}sert}, {Doyle}, {Goupy}, {Kramer}, {Leclercq}, {Martino}, {Mauskopf},
  {Mayet}, {Monfardini}, {Pajot}, {Pascale}, {Perotto}, {Pointecouteau},
  {Ponthieu}, {Rev{\'e}ret}, {Rodriguez}, {Savini}, {Schuster}, {Sievers},
  {Tucker}, \& {Zylka}}]{Adam2013}
{Adam}, R., {Comis}, B., {Mac{\'{\i}}as-P{\'e}rez}, J.~F., {et~al.} 2013, ArXiv
  e-prints

\bibitem[{{Allen} {et~al.}(2002){Allen}, {Schmidt}, \& {Fabian}}]{Allen2002}
{Allen}, S.~W., {Schmidt}, R.~W., \& {Fabian}, A.~C. 2002, \mnras, 335, 256

\bibitem[{{Arnaud} {et~al.}(2010){Arnaud}, {Pratt}, {Piffaretti},
  {B{\"o}hringer}, {Croston}, \& {Pointecouteau}}]{Arnaud2010}
{Arnaud}, M., {Pratt}, G.~W., {Piffaretti}, R., {et~al.} 2010, \aap, 517, A92

\bibitem[{{Calvo} {et~al.}(2013){Calvo}, {Roesch}, {D{\'e}sert}, {Monfardini},
  {Benoit}, {Mauskopf}, {Ade}, {Boudou}, {Bourrion}, {Camus}, {Cruciani},
  {Doyle}, {Hoffmann}, {Leclercq}, {Macias-Perez}, {Ponthieu}, {Schuster},
  {Tucker}, \& {Vescovi}}]{Calvo2013}
{Calvo}, M., {Roesch}, M., {D{\'e}sert}, F.-X., {et~al.} 2013, \aap, 551, L12

\bibitem[{{Cavagnolo} {et~al.}(2009){Cavagnolo}, {Donahue}, {Voit}, \&
  {Sun}}]{Cavagnolo2009}
{Cavagnolo}, K.~W., {Donahue}, M., {Voit}, G.~M., \& {Sun}, M. 2009, \apjs,
  182, 12

\bibitem[{{Cohen} \& {Kneib}(2002)}]{Cohen2002}
{Cohen}, J.~G. \& {Kneib}, J.-P. 2002, \apj, 573, 524

\bibitem[{{Comis} {et~al.}(2011){Comis}, {de Petris}, {Conte}, {Lamagna}, \&
  {de Gregori}}]{Comis2011}
{Comis}, B., {de Petris}, M., {Conte}, A., {Lamagna}, L., \& {de Gregori}, S.
  2011, \mnras, 418, 1089

\bibitem[{{Hasselfield} \& {ACT Collaboration}(2013)}]{Hasselfield}
{Hasselfield}, M. \& {ACT Collaboration}. 2013, in American Astronomical
  Society Meeting Abstracts, Vol. 221, American Astronomical Society Meeting
  Abstracts, 124.05

\bibitem[{{Komatsu} {et~al.}(1999){Komatsu}, {Kitayama}, {Suto}, {Hattori},
  {Kawabe}, {Matsuo}, {Schindler}, \& {Yoshikawa}}]{Komatsu1999}
{Komatsu}, E., {Kitayama}, T., {Suto}, Y., {et~al.} 1999, \apjl, 516, L1

\bibitem[{{Korngut} {et~al.}(2011){Korngut}, {Dicker}, {Reese}, {Mason},
  {Devlin}, {Mroczkowski}, {Sarazin}, {Sun}, \& {Sievers}}]{Korngut2011}
{Korngut}, P.~M., {Dicker}, S.~R., {Reese}, E.~D., {et~al.} 2011, \apj, 734, 10

\bibitem[{{Miranda} {et~al.}(2008){Miranda}, {Sereno}, {de Filippis}, \&
  {Paolillo}}]{Miranda2008}
{Miranda}, M., {Sereno}, M., {de Filippis}, E., \& {Paolillo}, M. 2008, \mnras,
  385, 511

\bibitem[{{Monfardini} {et~al.}(2011){Monfardini}, {Benoit}, {Bideaud},
  {Swenson}, {Cruciani}, {Camus}, {Hoffmann}, {D{\'e}sert}, {Doyle}, {Ade},
  {Mauskopf}, {Tucker}, {Roesch}, {Leclercq}, {Schuster}, {Endo}, {Baryshev},
  {Baselmans}, {Ferrari}, {Yates}, {Bourrion}, {Macias-Perez}, {Vescovi},
  {Calvo}, \& {Giordano}}]{Monfardini2011}
{Monfardini}, A., {Benoit}, A., {Bideaud}, A., {et~al.} 2011, \apjs, 194, 24

\bibitem[{{Monfardini} {et~al.}(2010){Monfardini}, {Swenson}, {Bideaud},
  {D{\'e}sert}, {Yates}, {Benoit}, {Baryshev}, {Baselmans}, {Doyle}, {Klein},
  {Roesch}, {Tucker}, {Ade}, {Calvo}, {Camus}, {Giordano}, {Guesten},
  {Hoffmann}, {Leclercq}, {Mauskopf}, \& {Schuster}}]{Monfardini2010}
{Monfardini}, A., {Swenson}, L.~J., {Bideaud}, A., {et~al.} 2010, \aap, 521,
  A29

\bibitem[{{Nagai} {et~al.}(2007){Nagai}, {Vikhlinin}, \& {Kravtsov}}]{gNFW}
{Nagai}, D., {Vikhlinin}, A., \& {Kravtsov}, A.~V. 2007, \apj, 655, 98

\bibitem[{{NIKA Collaboration}(2013)}]{main_run5}
{NIKA Collaboration}. 2013, in prep.

\bibitem[{{Plagge} {et~al.}(2012){Plagge}, {Marrone}, {Abdulla}, {Bonamente},
  {Carlstrom}, {Gralla}, {Greer}, {Joy}, {Lamb}, {Leitch}, {Mantz}, {Muchovej},
  \& {Woody}}]{Plagge2012}
{Plagge}, T.~J., {Marrone}, D.~P., {Abdulla}, Z., {et~al.} 2012, ArXiv e-prints

\bibitem[{{Planck Collaboration} {et~al.}(2013){Planck Collaboration}, {Ade},
  {Aghanim}, {Armitage-Caplan}, {Arnaud}, {Ashdown}, {Atrio-Barandela},
  {Aumont}, {Aussel}, {Baccigalupi}, \& et~al.}]{PSZ_cat}
{Planck Collaboration}, {Ade}, P.~A.~R., {Aghanim}, N., {et~al.} 2013, ArXiv
  e-prints

\bibitem[{{Pointecouteau} {et~al.}(1999){Pointecouteau}, {Giard}, {Benoit},
  {D{\'e}sert}, {Aghanim}, {Coron}, {Lamarre}, \&
  {Delabrouille}}]{Pointecouteau1999}
{Pointecouteau}, E., {Giard}, M., {Benoit}, A., {et~al.} 1999, \apjl, 519, L115

\bibitem[{{Pointecouteau} {et~al.}(2001){Pointecouteau}, {Giard}, {Benoit},
  {D{\'e}sert}, {Bernard}, {Coron}, \& {Lamarre}}]{Pointecouteau2001}
{Pointecouteau}, E., {Giard}, M., {Benoit}, A., {et~al.} 2001, \apj, 552, 42

\bibitem[{{Reichardt} {et~al.}(2013){Reichardt}, {Stalder}, {Bleem}, {Montroy},
  {Aird}, {Andersson}, {Armstrong}, {Ashby}, {Bautz}, {Bayliss}, {Bazin},
  {Benson}, {Brodwin}, {Carlstrom}, {Chang}, {Cho}, {Clocchiatti}, {Crawford},
  {Crites}, {de Haan}, {Desai}, {Dobbs}, {Dudley}, {Foley}, {Forman}, {George},
  {Gladders}, {Gonzalez}, {Halverson}, {Harrington}, {High}, {Holder},
  {Holzapfel}, {Hoover}, {Hrubes}, {Jones}, {Joy}, {Keisler}, {Knox}, {Lee},
  {Leitch}, {Liu}, {Lueker}, {Luong-Van}, {Mantz}, {Marrone}, {McDonald},
  {McMahon}, {Mehl}, {Meyer}, {Mocanu}, {Mohr}, {Murray}, {Natoli}, {Padin},
  {Plagge}, {Pryke}, {Rest}, {Ruel}, {Ruhl}, {Saliwanchik}, {Saro}, {Sayre},
  {Schaffer}, {Shaw}, {Shirokoff}, {Song}, {Spieler}, {Staniszewski}, {Stark},
  {Story}, {Stubbs}, {{\v S}uhada}, {van Engelen}, {Vanderlinde}, {Vieira},
  {Vikhlinin}, {Williamson}, {Zahn}, \& {Zenteno}}]{SPT_cat}
{Reichardt}, C.~L., {Stalder}, B., {Bleem}, L.~E., {et~al.} 2013, \apj, 763,
  127

\bibitem[{{Schindler} {et~al.}(1997){Schindler}, {Hattori}, {Neumann}, \&
  {Boehringer}}]{Schindler1997}
{Schindler}, S., {Hattori}, M., {Neumann}, D.~M., \& {Boehringer}, H. 1997,
  \aap, 317, 646

\bibitem[{{Staguhn} {et~al.}(2008){Staguhn}, {Allen}, {Benford}, {Sharp},
  {Ames}, {Arendt}, {Chuss}, {Dwek}, {Kovacs}, {Maher}, {Marx}, {Miller},
  {Moseley}, {Navarro}, {Sievers}, {Voellmer}, \& {Wollack}}]{Staguhn2008}
{Staguhn}, J., {Allen}, C., {Benford}, D., {et~al.} 2008, Journal of Low
  Temperature Physics, 151, 709

\bibitem[{{Sunyaev} \& {Zel'dovich}(1972)}]{SunyaevZeldovich1}
{Sunyaev}, R.~A. \& {Zel'dovich}, Y.~B. 1972, \apspr, 4, 173

\bibitem[{{Sunyaev} \& {Zel'dovich}(1980)}]{SunyaevZeldovich2}
{Sunyaev}, R.~A. \& {Zel'dovich}, Y.~B. 1980, \araa, 18, 537

\end{thebibliography}

\end{document}